# In Silico Prediction and Validation of LmGt Inhibitors Using QSAR and Molecular Docking Approaches

Pronama Biswas
Department of Biological Sciences
Dayananda Sagar University
Bengaluru, India
0000-0003-2950-1530

Madhavi Bhatt
Department of Biological Sciences
Dayananda Sagar University
Bengaluru, India
0009-0005-8946-8674

Belaguppa Manjunath Ashwin Desai
Department of ECE
Dayananda Sagar University
Bengaluru, India
0000-0002-4144-626

*Abstract*— Leishmaniasis caused by *Leishmania mexicana* relies on *Leishmania mexicana* gluscose transporter (LmGT) receptors, which play an important role in glucose and ribose uptake at different stages of parasite's life cycle. Previous efforts to identify LmGT inhibitors have been primarily based on in vitro screening. However, this conventional method is limited by inefficiency, high cost, and lack of specificity which leaves a significant gap in the development of targeted therapeutic candidates for LmGT. This study employs computational techniques to address this gap by developing a quantitative structure analysis relationship model, utilizing a support vector machine classifier to identify novel LmGt inhibitor. The QSAR model achieved an accuracy of 0.81 in differentiating active compounds. Molecular docking further validated the identified inhibitors, revealing strong binding affinities with a top score of -9.46. The docking analysis showed that the inhibitors formed multiple hydrogen bonds and occupied the same binding pockets as Phase 3 drug candidate. The tested inhibitors were derived from natural sources, which suggest reduced side effects and improved biocompability. This combined approach demonstrates the power of computational models in accelerating drug discovery, with implication for more efficient and biocompatible therapies against *Leishmania mexicana*.

## I. INTRODUCTION

Leishmaniasis caused by twenty different species of protozoan parasite *Leishmania* from the *trypanosomatid* family and transmitted through the bite of an infected sandflies. It is known as a neglected tropical disease (NTD) and impats over a billion people per year in Europe, America, Asia, and Africa. Leishmaniasis is responsible for causing distinct clinical forms such as VL (visceral leishmaniasis), CL (cutaneous leishmaniasis), DCL (diffuse cutaneous leishmaniasis), MCL (mucocutaneous leishmaniasis), affecting various hosts species.

*L. mexicana*, a new world species, is the primary cause for CL, where single or multiple skin lesions gradually develop into ulcerated sores. Occasionally, *L. mexicana* leads to DCL, in which lesions spread across skin and harder to treat. These lesions can lead to secondary bacterial infection as well as MCL, causing severe damage to the mucosa of the nose, throat, and mouth. Currently, no approved vaccine available for humans, and available drug treatments exhibit significant toxicity and increasing drug resistance.

*L. mexicana* depends on nutrition acquisition for its survival and pathogenicity, mainly glucose and ribose in the absence of glucose. LmGT1, LmGT2, LmGT3 are three glucose transport isoforms belongs to *L. mexicana* glucose transport family (LmGT). These isoforms facilitate parasite's metabolic adaptation through glucose uptake. Among these three isoforms, LmGT2 is essentially responsible for glucose uptake and shows a higher affinity for ribose transport than LmGT1 and LmGT3 [3,4].

This research aimed to construct a computational framework to distinguish between active and inactive molecules targeting LmGT2 by incorporating QSAR technique followed by molecular docking, two widely utilized computational approaches in drug discovery [5]. The model was designed to accurately predict potential inhibitors, classifying them based on their likelihood of inhibiting glucose transport in *L. mexicana*. The identified active compounds were then validated using molecular docking, assessing their binding affinity and interactions with the LmGT2 binding sites. Finally, the docking results of these compounds were evaluated against those of a Phase 3 drug candidate, offering insights into their comparative efficacy and potential as therapeutic agents. By leveraging these computational techniques, this study offers a streamlined approach to uncover potential inhibitors for blocking glucose uptake and impair parasite viability. By applying these computational methods, this study provides a streamlined approach for discovering potential inhibitors that disrupt glucose uptake and impair parasite viability

## II. METHODS

### A. Data Collection and Cleaning

The ChEMBL database was used to collect data for finding possible LmGT inhibitors and their inhibitory activities. The target with ChEMBL ID "CHEMBL3431938," was selected due to large IC50 dataset from *Leishmania*, covering many compounds and assay conditions. The dataframe "Rawdata_LmGT.csv" was generated, which contained only IC50 values, resulted in 792 molecules. For data cleaning, entries without IC50 values and missing SMILES notations were removed, IC50 values were standardized to nanomolar (nM), and compound redundancy was checked. Compound with high IC50 standard deviation was classified as redundant and removed. After cleaning, 791 molecules remained in the final dataset, saved as "Cleaned_data_LmGt.csv."

### B. Descriptors and Fingerprint Calculation

Using RDKit (https://www.rdkit.org/), 791 molecular descriptors were calculated from the SMILES strings to capture key physicochemical and structural properties. Invalid molecules were identified and excluded from the dataset, and the final set of descriptors was saved as "RDKit_Descriptors_LmGT.csv." Additionally, various molecular fingerprints, such as Morgan, Avalon, MACCS keys, topological torsion, and atom pair fingerprints, were generated to represent structural features. These fingerprints were combined and saved as "RDKit_FPs_LmGT.csv" for subsequent modeling. Altogether, the descriptors and fingerprints provided a comprehensive set of 2,937 features for QSAR modelling.

*C. Data Preprocessing, Feature Selection and Scaling*

Compounds were evaluated based on in vitro biological activity using IC50 values, which represent the half-maximal concentration of the drug. IC50 values were converted to pIC50 and binarized to classify compounds as active or inactive, using a threshold of 10 μM (corresponding to pIC50 ≥ 5) for active compounds and pIC50 < 5 for inactive ones. The compounds with IC50 ≤ 10 μM indicate potent and effective inhibitors [6]. This aligns with pharmacological standards for identifying potential lead therapeutic candidates. The features were thoroughly checked for missing or infinite values. The final dataset consisted of 382 active compounds (Class 1) and 409 inactive compounds (Class 0). To enhance model performance, a variance threshold was applied to remove low-variance features, followed by mutual information-based filtering to select the most relevant features. 989 features were retained for QSAR modelling, which were scaled. The preprocessed data was saved as "Classification_data_LmGT.csv."

*D. Model Training and Testing*

Several machine learning models were explored such as Logistic Regression, Random Forest, and Support Vector Machine (SVM), were trained using a stratified 70:30 train-test split, with balanced class weights applied to address potential class imbalance. Among the models tested, SVM demonstrated the best performance and was further fine-tuned using GridSearchCV with hyperparameters 'C=31.27', a polynomial kernel ('degree=3'), and gamma=auto. The polynomial kernel was selected to handle nonlinearly separation data by mapping it to a higher dimensional space, with the degree controlling the complexity of the decision boundary. To enhance the model's performance, 5-fold cross-validation was employed, where the dataset was split into five subsets, and the model was iteratively trained on four subsets while tested on the fifth [7]. This approach evaluated performance across multiple metrics such as accuracy, precision, recall, and F1-score. Additionally, the model's generalizability was verified by calculating the Area Under the Curve (AUC) for the Receiver Operating Characteristic (ROC) curve in each fold.

*E. Tools and Libraries*

This study was conducted using Python, utilizing RDKit for descriptor and fingerprint generation, Scikit-learn for model development (https://scikit-learn.org), and Matplotlib for visualization (https://matplotlib.org/). Computational tasks were executed on a high-efficient server featuring an AMD EPYC 7742 64-core processor, 503 GiB of RAM, and NVIDIA A100 GPUs running on Ubuntu 22.04.4. All processes were carried out within a dedicated Conda environment.

*F. Dataset for prediction*

To identify novel LmGT inhibitors, a dataset comprising exclusively of Genuine Natural (GN) and Naturally Sourced (NS) compounds was obtained from the MolPort database (https://www.molport.com). GN compounds naturally occur, though they may also be synthetically produced, while NS compounds are directly isolated from natural sources. This dataset was evaluated using the trained SVM classifier to determine the inhibitory potential of these compounds against LmGT. The approach enabled the identification of potential novel inhibitors with significant biological relevance. Furthermore, due to the natural origin of GN and NS compounds, they are expected to exhibit reduced toxicity and enhanced biocompatibility.

*G. Molecular Docking*

The top 10 predicted inhibitors, showing the highest probabilities of being active, were chosen for further validation through molecular docking protocol [8]. These molecules were docked against the LmGT2 target, and their binding affinities were compared to that of a known inhibitor, Cytidine (CHEMBL95606) [9]. The protein structure of LmGT was retrieved as a Protein Data Bank (PDB) file, which was obtained from AlphaFold when searched with the UniProt ID "O61059". The protein quality was tested using PROCHECK (https://saves.mbi.ucla.edu/) and VoroMQA (https://bioinformatics.lt/wtsam/voromqa) websites. The stereochemistry of the SMILES was corrected and converted to SDF files, followed by conversion to PDBQT files with Gasteiger charges added using a command from Open Babel. LmGT2 and Cytidine were prepared using AutoDockTools and saved in PDBQT format. Molecular docking of Cytidine and LmGT2 was performed using a Vina Script-based method in a Conda environment using a server, which was accessed through Visual Studio Code. The molecular docking of the predicted molecules was performed in a batch using the same script-based method. Molecular docking was performed in triplicates, and binding affinity values were assessed by calculating their average and standard deviation. The protein-ligand docked complex was analyzed to visualize amino acids and binding pockets using UCSF Chimera and Discovery Studio 2021 Client. Fig. 1 summarises the methodology implemented in this study.

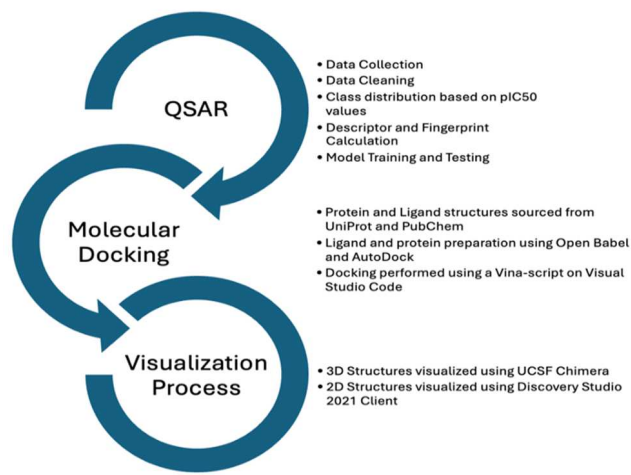

Fig. 1. Schematic representation of the QSAR, molecular docking and visualization methodology involved in the study.

### III. RESULTS AND DISCUSSIONS

*A. Dataset Characteristics*

After the data cleaning and preprocessing steps, a total of 791 molecules were left in the dataset, among which 409 were inactive molecules (pIC50 < 5) and 382 were active (pIC50 ≥ 5). Fig. 2. illustrates the distribution of pIC50 values across both active and inactive molecules. As depicted, the majority of inactive compounds are concentrated around a pIC50 value of 5, while active compounds are distributed between pIC50 values of 5 and 7.

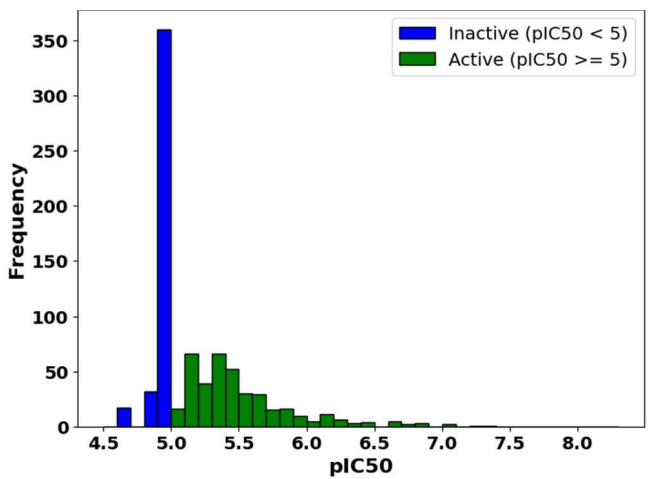

Fig. 2. Distribution of pIC50 values for active (pIC50 ≥ 5) and inactive (pIC50 < 5) molecules in the dataset after data cleaning. The dataset contains 409 inactive and 382 active molecules, with a clear concentration of inactive compounds around pIC50 = 5 and active compounds between pIC50 = 5 and 7.4

### B. Performance of the model

The SVM model, fine-tuned with specific hyperparameters, {'C': 31.27, 'kernel': Poly, and 'degree': 3, 'gamma': auto}, demonstrated robust performance in predicting LmGT2 inhibitors across multiple cross-validation folds. As shown in Table I, the ROC AUC was obtained at 0.86, with the model achieving an average accuracy of 0.81. It consistently maintained high precision (0.79) and recall (0.82) for active compounds, indicating a well-balanced trade-off between precision and recall. The average F1 score was 0.81 for inactive compounds and 0.80 for active compounds. The ROC curves for each validation fold (Fig. 3) demonstrated the classifier's strong discriminatory power, with the mean ROC curve indicating reliable performance across thresholds. This highlights the SVM model's effectiveness not only in accurately predicting LmGT2 inhibition but also in providing consistent results, making it suitable for both broad screening and detailed pharmacological evaluation.

TABLE I. SUMMARY OF AVERAGE METRICS FOR THE SVM CLASSIFIER FOR FIVE FOLDS

| Metric | Value |
|---|---|
| Average Accuracy | 0.81 |
| Average Precision (Class 0) | 0.82 |
| Average Recall (Class 0) | 0.80 |
| Average F1-Score (Class 0) | 0.81 |
| Average Precision (Class 1) | 0.79 |
| Average Recall (Class 1) | 0.82 |
| Average F1-Score (Class 1) | 0.80 |
| Average ROC AUC | 0.86 |

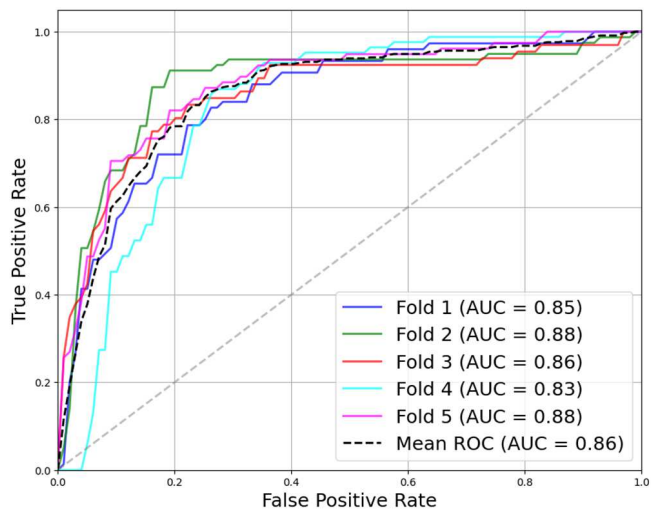

Fig. 3. ROC Curves for each fold in Cross-Validation, illustrating the model's discriminative performance across the folds with AUC values ranging between 0.83 to 0.88.

### C. Docking of predicted molecules

Out of the 10,290 inhibitors tested, the trained model predicted 1721 as active. These predicted active compounds were screened using Pan Assay Interference Compounds (PAINS) and BRENK filtering, reducing the dataset to 288 molecules by removing potential false positives or compounds likely to interfere in biological assays [10]. Binding affinity, represented by the Gibbs free energy of binding (ΔG), quantifies the strength of interaction between a protein and a ligand, with more negative values indicating stronger binding. Docking of the top 10 predicted inhibitors against LmGT2 revealed binding affinities that were generally stronger than that of the known inhibitor, cytidine, which exhibited a binding affinity of -7.47 ± 0.751 (Table II). For example, molecule 7 (MolPort-005-945-897) and molecule 10 (MolPort-001-742-665) exhibited high binding affinity values of -9.46 ± 0.27 kcal/mol and -9.25 ± 0.01 kcal/mol, respectively. These results highlighted that these compounds might have a potentially stronger inhibitory effect, making them promising alternatives than cytidine. The minimal standard deviation demonstrated the reproducibility of the docking protocol.

3D visualization demonstrated that the binding pockets of both cytidine and the predicted inhibitors were identical (Fig. 4a, c, and f), suggesting that they all bind to the active site of LmGT2. Furthermore, 2D visualization showed that the docked complexes of cytidine, molecule 7, and molecule 4 shared key binding residues, Asp 509, Gln 508, Leu 370, Ile 366, and Ser 369, which are highlighted in red circles (Fig. 4b, d, and e). These observations indicate that the predicted therapeutic candidates target the active site of LmGT2, due to presence of shared binding residues with Cytidine. This supports their potential as potent inhibitors with possibly greater efficacy. Additionally, identifying key interaction sites provides valuable insights for optimizing these compounds, facilitating structure-based drug design and guiding future experimental validation studies.

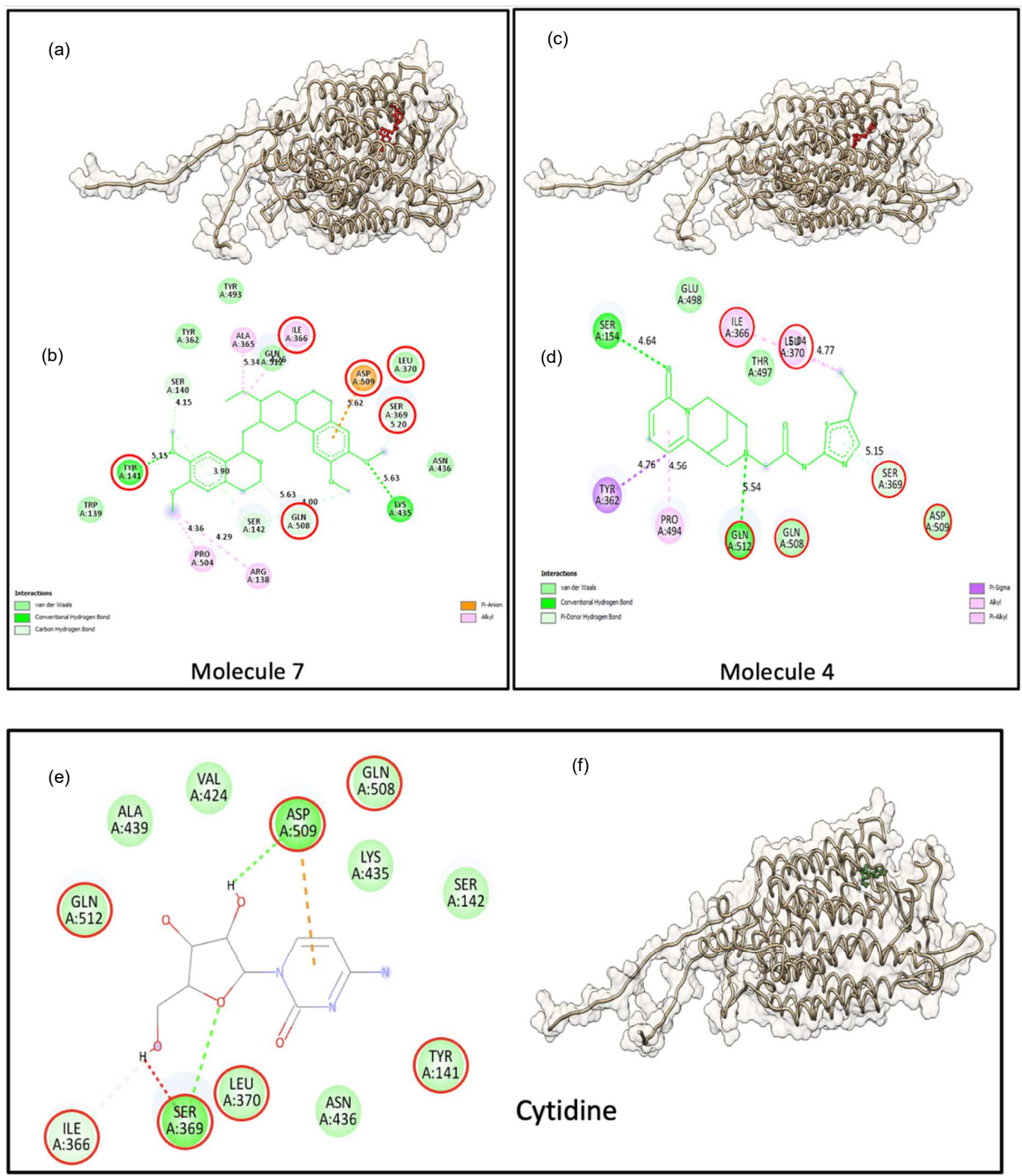

Fig. 4. Visual representation of binding pockets and binding amino acids of the docked complex. (a) 3D-visualisation of molecule 7 (MolPort-005-945-897), (b) 2D visualisation of molecule 7 (MolPort-005-945-897), (c) 3D-visualisation of molecule 4 (MolPort-007-980-947), (d) 2D visualisation of molecule 4 (MolPort-007-980-947), (e) 2D-visualisation of Cytidine and (f) 3D-visualisation of Cytidine.

TABLE II. BINDING AFFINITY VALUES OF TOP 10 PREDICTED INHIBITORS

| Sl. no | Prediction Probability with respect to class 0 | MOLPORT ID | Binding affinity (kcal/mol) |
|---|---|---|---|
| 1 | 0.0000003 | MolPort-039-055-596 | -8.65 ± 0.27 |
| 2 | 0.026 | MolPort-038-386-535 | -9.18 ± 0.36 |
| 3 | 0.029 | MolPort-003-666-755 | -9.10 ± 0.44 |
| 4 | 0.032 | MolPort-007-980-947 | -8.25 ± 0.08 |
| 5 | 0.036 | MolPort-038-386-534 | -9.03 ± 0.40 |
| 6 | 0.042 | MolPort-044-560-304 | -8.17 ± 0.01 |
| 7 | 0.084 | MolPort-005-945-897 | -9.46 ± 0.27 |
| 8 | 0.066 | MolPort-028-610-178 | -8.61 ± 0.04 |
| 9 | 0.112 | MolPort-009-683-217 | -8.29 ± 0.13 |
| 10 | 0.115 | MolPort-001-742-665 | -9.25 ± 0.01 |

CONCLUSIONS

This study highlights the effective application of QSAR modeling and molecular docking to predict ten promising LmGT2 inhibitors. After thorough data cleaning, a dataset of 791 molecules (382 active, 409 inactive) were analyzed. The SVM Classifier, optimized with specific hyperparameters, achieved an accuracy of 0.81 along with a ROC AUC score of 0.86, exhibiting strong predictive power of identifying LmGT2 inhibitors, with a precision of 0.79 and recall of 0.82 for active compounds. Molecular docking results for the top two predicted inhibitors, along with the known inhibitor cytidine, revealed that Molecule 7 (MolPort-005-945-897) exhibited the strongest binding affinity of -9.46 ± 0.27, followed by Molecule 10 (Molport-001-742-665) with a binding affinity of -9.25 ± 0.01, both surpassing the affinity of cytidine. These inhibitors also shared common binding sites with cytidine, confirming their alignment at the same active site. The integration of QSAR modeling and molecular docking could succesfully identify high-affinity LmGT inhibitors, offering likely candidates for further experimental exploration.

ACKNOWLEDGEMENT

The authors thank AIC-DSU for supporting them with the server for conducting this study.

REFERENCES

[1] G.J. Wijnant, F. Dumetz, L. Dirkx, D. Bulte, B. Cuypers, K. van Bocxlaer, and S. Hendrickx, "Tackling drug resistance and Other causes of Treatment Failure in Leishmaniasis," Front. Trop. Dis., vol. 3, May 2022, article 837460. https://doi.org/10.3389/fitd.2022.837460

[2] I. Abadias-Granado, A. Diago, P. A. Cerro, A. M. palma-Ruiz, and Y. Gilaberte, "Cutaneous and mucotaneous leishmaniasis," Actas Dermo-Sifiliograficas, vol. 112, no. 7, pp. 601-618, Jul-Aug 2021. https://doi.org/10.1016/j.ad.2021.02.008

[3] C. M. Naula, F. M. logan, P. E. Wong, M. P. Barrett, and R. J. burchmore, "Definition of residues that confer substrate specificity in a sugar transport," J. Biol. Chem., vol. 285, no. 39, pp.29721-29728, Sept. 24, 2010. https://doi.org/10.1074/jbc.M110.106815

[4] R. J. S. Burchmore, D. Rodriguez-Contreras, K. McBride, M. P. Barrett, G. Modi, D. Sacks, and S. M. Landfear, "Genetic characterization of glucose transporter function in Leishmania mexicana," Proc. Natl. Acad. Sci. USA, vol. 100, no. 7, pp. 3901-3906, Apr. 1, 2003. https://doi.org/10.1073/pnas.0630165100

[5] B. J. Neves, R. C. Braga, C.C. Melo-Filho, J.T. Moreira-Filho, E.N. Muratov, and C.H. Andrade, "QSAR-Based Virtual Screening: Advances and Applications in Drug Discovery," Frontiers in Pharmacology, vol. 9, 2018, Art No. 1275. https://doi.org/10.3389/fphar.2018.01275

[6] G.V. Paolini, R.A. Lyons, and P. laflin, "How desirable are your IC50s? A way to enhance screening-based decision making," Journal of Biomolecular Screening, vol. 15, no. 10, pp. 1183-1193, 2010. https://doi.org/10.1177/1087057110384402

[7] I. Shadeed, J. Alwan, and D. Abd, "The effect of gamma value on support vector machine performance with different kernals," international Journal of Electrical and Computer Engineering (IJECE), vol. 10, pp. 5497-5506, 2020. https://doi.org/10.11591/ijece.v10i5.pp5497-5506

[8] S. Forli, R. Huey, M.E. Pique, M.F. Sanner, D.S. Goodsell, and A.J. Olson, "Computational protein–ligand docking and virtual drug screening with the AutoDock suite," Nat Protoc, vol. 11, pp. 905–919, 2016. https://doi.org/10.1038/nprot.2016.051.

[9] S. Bhattacharya, T. Bhattacharyya, S. Khanra, R. Banerjee, and J. Dash, "Nucleoside Derived mettallohydrogel Includes Cell Death in Leishmania Parasites," ACS Infectious Diseases, vol. 9, no. 9, pp. 1676-1684, Sep. 2023. https://doi.org/10.1021/ascinfecdis.2c00635

[10] J. B. Baell and G. A. Holloway, "New Substructure Filters for Removal of Pan Assay Interference Compounds (PAINS) from Screening Libraries and for Their Exclusion in Bioassays," Journal of medicinal Chemistry, vol. 53, pp. 2719-2740, 2010. https://doi.org/10.1021/jm901137j